\documentclass[conference]{IEEEtran}
\IEEEoverridecommandlockouts

\usepackage{cite}
\usepackage{amsmath,amssymb,amsfonts}
\usepackage{algorithmic}
\usepackage{graphicx}
\usepackage{textcomp}
\usepackage{xcolor}
\usepackage{xspace}
\usepackage{tcolorbox}
\usepackage{color, colortbl}
\newcommand{\etal}{\emph{et al}.\@\xspace}
\newcommand*{\eg}{\emph{e.g.},\@\xspace}
\newcommand*{\ie}{\emph{i.e.},\@\xspace}

\def\BibTeX{{\rm B\kern-.05em{\sc i\kern-.025em b}\kern-.08em
 T\kern-.1667em\lower.7ex\hbox{E}\kern-.125emX}}
\definecolor{Gray}{gray}{0.9}

\begin{document}

\title{Predicting Flaky Tests Categories using Few-Shot Learning}
\makeatletter
\newcommand{\linebreakand}{%
  \end{@IEEEauthorhalign}
  \hfill\mbox{}\par
  \mbox{}\hfill\begin{@IEEEauthorhalign}
}
\makeatother
% \author{\IEEEauthorblockN{Anonymous Authors}}
\author{
\IEEEauthorblockN{Amal Akli}
\IEEEauthorblockA{
Ecole Nationale Supérieure d'Informatique \\
Algeria \\
ha\_akli@esi.dz}
\and
\IEEEauthorblockN{Guillaume Haben}
\IEEEauthorblockA{
University of Luxembourg \\
Luxembourg \\
guillaume.haben@uni.lu}
\and
\IEEEauthorblockN{Sarra Habchi}
\IEEEauthorblockA{
Ubisoft \\
Canada \\
sarra.habchi@ubisoft.com}
\linebreakand
\IEEEauthorblockN{Mike Papadakis}
\IEEEauthorblockA{
University of Luxembourg \\
Luxembourg \\
michail.papadakis@uni.lu}
\and
\IEEEauthorblockN{Yves Le Traon}
\IEEEauthorblockA{
University of Luxembourg \\
Luxembourg \\
yves.letraon@uni.lu}
}

\maketitle

\begin{abstract}

Flaky tests are tests that yield different outcomes when run on the same version of a program. This non-deterministic behaviour plagues continuous integration with false signals, wasting developers' time and reducing their trust in test suites. Studies highlighted the importance of keeping tests flakiness-free. Recently, the research community has been pushing forward the detection of flaky tests by suggesting many static and dynamic approaches. While promising, those approaches mainly focus on classifying tests as flaky or not and, even when high performances are reported, it remains challenging to understand the cause of flakiness. This part is crucial for researchers and developers that aim to fix it. To help with the comprehension of a given flaky test, we propose FlakyCat, the first approach for classifying flaky tests based on their root cause category. 
FlakyCat relies on CodeBERT for code representation and leverages a Siamese network-based Few-Shot learning method to train a multi-class classifier with few data. 
We train and evaluate FlakyCat on a set of 343 flaky tests collected from open-source Java projects. 
%To train and evaluate this approach, we %To this end, we gathered a dataset of 343 flaky tests with their corresponding categories. 
%We compare our approach with the existing machine learning approaches adapted to the multi-class classification problem. 
Our evaluation shows that FlakyCat categorises flaky tests accurately, with a weighted F1 score of 70\%. Furthermore, we investigate the performance of our approach for each category, revealing that \textit{Async waits}, \textit{Unordered collections} and \textit{Time}-related flaky tests are accurately classified, while \textit{Concurrency}-related flaky tests are more challenging to predict. Finally, to facilitate the comprehension of FlakyCat's predictions, we present a new technique for CodeBERT-based model interpretability that highlights code statements influencing the categorization.

\end{abstract}


\section{Introduction}

Continuous Integration (CI) plays a key role in nowadays' software development life cycle~\cite{shahin2017continuous, CI}. 
CI ensures the quick application of changes to a main code base by automatically running a variety of tasks. Those changes can be responsible for building the program and its dependencies, performing checks (\eg static analysis), and running test suites to maintain code integrity and correctness. An important assumption for practitioners is that tasks are deterministic, \ie regardless of the context of the execution of a same task, results need to be persistent. 

Unfortunately, in practice, this is not always the case. Previous research has identified test flakiness as one of the main issues in the application of automated software testing~\cite{Micco2017,memon2017taming}. A flaky test is a test that passes and fails when executed on the same version of a program. Flakiness hinders CI cycles and prevents automatic builds due to false signals, resulting in undesirable delays. Furthermore, surveys~\cite{habchi2021qualitative, Eck2019, gruber2022survey} show that flakiness affects developers' productivity as they spend a considerable time and effort investigating the nature and causes of flaky tests. 

To alleviate this issue, researchers have proposed tools that help detect flaky tests. In particular, IDFlakies~\cite{Lam2019iDFlakies} and Shaker~\cite{Silva2020} detect flakiness in test suites by running tests in different setups. However, rerunning tests, especially for a large number of times, is resource-intensive and might not be a scalable solution. For this reason, researchers recently suggested alternative approaches to detect flaky tests based on features that do not require any test execution \cite{King2018,pinto2020vocabulary,camara2021use}. Although promising, these approaches mainly focus on classifying tests as flaky or not without any additional explanation. Unfortunately, the absence of any additional information about the cause of flakiness is insufficient to comprehend and investigate the cause of the (flaky) failures. This means that additional investigation is required to understand the nature of flakiness and identify the culprit code elements that need to be fixed~\cite{Eck2019}.

Another important line of research in the area regards automated approaches that aim at helping to locate the root causes and suggest potential flakiness fixes~\cite{De-Flake,Lam2019RootCausing,flakyloc}. However, research on automatically fixing flakiness is still at an early stage: tools often focus on one category of flakiness and with few examples. For instance, iFixFlakies~\cite{Shi2019iFix} and ODRepair~\cite{li2022repairing} focus only on dealing with test order dependencies, which is one of the main causes of test flakiness. Flex~\cite{Dutta2020} automatically fixes flakiness due to algorithmic randomness in machine learning algorithms. 

We believe that both developers and researchers could benefit from additional information that could assist them in gaining a better understanding of flaky tests, once they have been detected. Therefore, we propose FlakyCat, a learning-based flakiness categorization approach that identifies the key reason/category of the test failures. FlakyCat relies on CodeBERT~\cite{feng2020codebert} to build appropriate code representations and allow static test flakiness predictions, i.e., predictions based solely on test code. 

Another limitation of previous work, relying on supervised learning, regards the need for large volumes of available data. Unfortunately, debugged flaky test data is scarce, inhibiting the application of learning-based methods. To deal with this issue we leverage the Few-Shot learning capabilities of Siamese networks, which we combine with the CodeBERT representations to learn flakiness categories from a limited set of data (flaky tests). 

To evaluate FlakyCat, we collect a set of 343 flaky tests from open-source Java projects and categorize them based on flakiness fixing commit messages, that are applied on both tests and projects' source code. We then use this annotated flaky test set to train and evaluate FlakyCat and compare it with some relevant baselines.

In particular, our empirical evaluation aims at answering the following three research questions:

\begin{itemize}
    \item \textsc{\textbf{RQ1:}} How effective is FlakyCat compared to approaches based on traditional supervised learning? \\
   % \textbf{Approach:} We compare FlakyCat with existing machine learning approaches adapted to the multi-class classification problem. \\
    \textbf{Findings:} Our results show that FlakyCat is capable of predicting flakiness categories with an F1 score of 70\%, outperforming classifiers based on traditional supervised machine learning that was used in previous work.

    \item \textsc{\textbf{RQ2:}} How effective is FlakyCat in predicting each one of the considered flakiness categories?\\
    %\textit{Motivation} We investigate more on the performance of our approach for each category. We wonder if the information embedded in the CodeBERT representation of test code can help classifying the different categories.\\
    \textbf{Findings:} FlakyCat classifies accurately flaky tests related to Async waits, Unordered collection, and Time, with an average F1 score of 70\%. However, the approach shows difficulty in classifying Concurrency-related flaky tests, since these cases are related to the interaction of threads and processes and are easily confused with Async waits. 

    \item \textsc{\textbf{RQ3:}} How do statements of the test code influence the predictions of FlakyCat?\\
    %\textit{Motivation} An important step in many machine learning tasks is model interpretability. We want to check what information the model learnt.\\
    \textbf{Findings:} We found that statements that lead the model to predict the categories of \textit{Time} and \textit{Unordered collections} tend to enclose information related to the corresponding flakiness cause. On another hand, it is more challenging to identify a correlation between the statements influencing the categorisation of the \textit{Async waits} and \textit{Concurrency} categories. 
\end{itemize}

In summary, our contributions can be summarised as follows:
 \begin{itemize}
    \item \textbf{Dataset} We collected 343 flaky tests alongside their category of test flakiness.
     \item \textbf{Model} We present FlakyCat, a new approach based on Few-Shot Learning and CodeBERT to classify flaky tests with regard to their flakiness category. 
     \item \textbf{Interpretability} We introduce a technique to explain what information is learnt by models using CodeBERT as code representation.
 \end{itemize}

To enable reproducibility and replicability of our work, we make the dataset used to evaluate FlakyCat and the scripts publicly available in our replication package \footnote{https://anonymous.4open.science/r/FlakyCat-5033/}.

The paper is organized as followed: Related works are presented in section~\ref{sec:related_works}. Section~\ref{sec:flakycat} presents how we designed and implemented FlakyCat. Section~\ref{sec:inter} introduces our interpretability technique. Section~\ref{sec:eval} describes how we collected our dataset and evaluated our study. Section~\ref{sec:results} presents the results of our study. We further discuss different use cases in Section~\ref{sec:discussion}. Finally, section~\ref{sec:threats} gives details about the threats to the validity of this study.

\section{Related Work}
\label{sec:related_works} %TODO: Emphasize the comparison with this work
% Starting research
Recently, practitioners from the industry reported struggling with flakiness and highlighted the need to find solutions to the problem\cite{Micco2017,JiangHuawei,Mozilla,Harman2018,FlakinessSpotify}. Consequently, researchers from academia started to draw their attention to the matter. Luo~\etal presented the first empirical study to understand and categorize the root causes of flakiness, they analyzed 201 flaky tests and identified 10 root causes of flakiness, the top ones being Asynchronous waits, concurrency, and test order dependency. Using the same taxonomy defined by Luo \cite{Luo2014}, Eck~\etal\cite{eck} classified 200 flaky tests and identified four new causes of flakiness. 
% Surveys
Over the years, several surveys were carried on to identify the sources, impacts and existing strategies to mitigate flakiness by interrogating developers and practitioners \cite{habchi2021qualitative, Eck2019, gruber2022survey, parry2022surveying}. Parry \etal presented the state of the art of academic research in another survey\cite{Parry2021}.
% Detection - Tools
Researchers presented different tools and approaches to detect flaky tests in a more efficient way. Notably, DeFlaker\cite{Bell2018}, IDFlakies\cite{Lam2019iDFlakies}, Shaker\cite{Silva2020} and NonDex\cite{gyori2016nondex} attempt to facilitate the detection of flaky tests compared to exhaustive reruns. 
% Detection - ML based
Because the cost of running tests is viewed as expensive, researchers also sought to suggest static alternatives for the detection. Different approaches relying on machine learning were introduced. Pinto \etal~\cite{Pinto2020} and following replication studies\cite{Haben2021,Camara2021VocabExtendedReplication} presented a vocabulary-based model using elements from the test code to classify tests as flaky or not. Others investigated the use of test smells~\cite{camara2021use} and code metrics~\cite{Pontillo} for predicting flaky tests. 
Trying to outperform the performances of existing approaches, others relied on a mix of static and dynamic features, like FlakeFlagger\cite{FlakeFlagger} or Flake16\cite{flake16}.
% Fixing 
Fixing flakiness is also an aspect that has recently been investigated. Shi \etal introduced iFixFlakies\cite{Shi2019iFix} to fix order-dependent flaky tests. At Google, Ziftci \etal suggested using coverage differences, between passing and failing executions of flaky tests to guide developers to understand the underlying problem. Coverage information is also used by FlakyLoc~\cite{flakyloc}, which leverages spectrum-based fault localization to locate the root cause of flakiness in web apps. Logs are also frequently considered to be a useful source of information in understanding root causes of flakiness ~\cite{Lam2019RootCausing}.
% Clother
Closer to our work, Flakify\cite{fatima2021flakify} used CodeBERT\cite{feng2020codebert} as a pre-trained language-based model for their predictor. As their goal is to classify tests as flaky or not, we are the first to focus on predicting the category of flakiness for each test. 
Few-Shot Learning is widely used in computer vision\cite{sun2021research}. In software engineering though, fewer studies used this approach for their task. Notably, studies suggested using this model for vulnerability detection~\cite{khajezade2022evaluating} and code clone detection~\cite{he2020vul}, but none were carried out for flakiness. About those pre-trained language models, Wan~\etal~\cite{wan2022they} investigated their ability to capture the syntax structure of source code, and report that they are efficient for code processing tasks. 

% Talk about What Do They Capture? - A Structural Analysis of Pre-Trained Language Models for Source Code\cite{wan2022they}

\section{FlakyCat}
\label{sec:flakycat}

\begin{figure*}[htbp]
\centerline{\includegraphics[width=0.7\textwidth,scale=1]{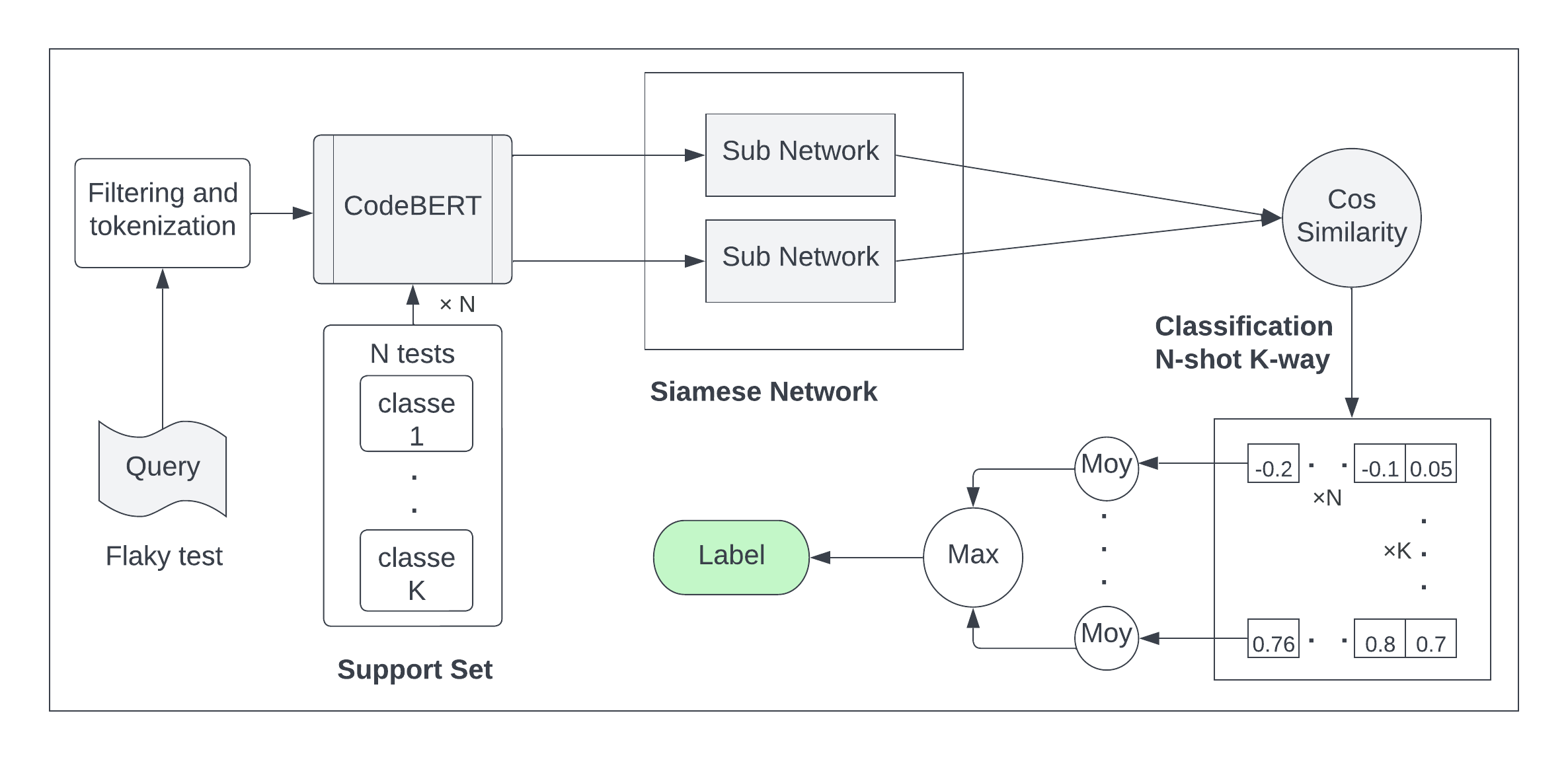}}
\caption{An overview of FlakyCat }
\label{fig:general_acrh}
\end{figure*}

In this section, we present the design and implementation of FlakyCat.
Figure~\ref{fig:general_acrh} presents an overview of the main steps of FlakyCat, code transformation and classification. 

\subsection{Step 1: Test transformation}

\subsubsection{Scope} We rely on the test code to assign flaky tests to different categories.
Previous studies showed that flakiness finds its root causes in the test in more than 70\% of the cases\cite{Luo2014,Lam}. 
Hence, focusing on the test code allows us to capture the nature of flakiness while minimizing the overall cost of FlakyCat. 
Indeed, considering the code under test would require running the tests and collecting the coverage, which entails additional requirements and costs.
%Then, we use Few-Shot Learning (FSL) and Siamese network for the classification. The general architecture of our solution is presented in Figure \ref{fig:general_acrh}. 

\subsubsection{Test vectorization}
In order to perform a source code classification task, we first need to transform the code into a suitable representation that will be fed to the classification model. Previous flakiness studies relied on predefined features~\cite{FlakeFlagger,King2018}, test smells~\cite{camara2021use,FlakeFlagger}, and code vocabulary \cite{pinto2020vocabulary,Haben2021,Camara2021VocabExtendedReplication} to transform code into vectors.
Test smells convey limited information, \ie smell presence, and may not be sufficient to capture the nature of flakiness. 
In the same vein, vocabulary-based approaches do not grasp the semantics of the code~\cite{chakraborty2021deep} and tend to overfit to the vocabulary present in specific projects~\cite{camara2021use,Haben2021}. 

Recently, code embeddings from pre-trained language models were also considered for source code representation~\cite{fatima2021flakify,zhou2021assessing}. Pre-trained language models allow the encoding of code semantics and are intended for general-purpose tasks such as code completion, code search, and code summarization.
Considering these benefits, we use the pre-trained language model CodeBERT \cite{feng-etal-2020-codebert} to generate source code embeddings. 
CodeBERT can learn the syntax and semantics of the code and doesn't require any predefined features \cite{wan2022they}. 
%\subsubsection{Why CodeBERT?}
%Many language models pre-trained on source code have been proposed, mainly from the BERT family. 
It supports both programming language and natural language. It has been developed with a multi-layer transformer architecture~\cite{transformer}, and trained on over six million pieces of code involving six programming languages (Java, Python, JavaScript, PHP, Ruby, and Go). 
In the following, we explain how we used CodeBERT to vectorize the test code.
%Similarly to CodeBERT \cite{feng-etal-2020-codebert}, GraphCodeBERT \cite{Guo2021GraphCodeBERTPC} is also pre-trained with six programming languages. In addition to the structure of the source code GraphCodeBERT includes information about the data flow.
% By only following the data flow of the test case code, this model cannot capture the type and value of a global variable as an example. 
%TreeBERT \cite{TreeBert} has been pre-trained using AST representations of source code in Java and Python. CuBERT \cite{cubert}, on the other hand, is pre-trained using only the Python source code. Unlike TreeBERT and CuBERT, CodeBERT is trained on six programming languages and gives us more choice in selecting our data, and constrained to the three models, CodeBERT requires no data transformation. Furthermore, since the length of inputs in these models is set to a maximum of 512, CodeBERT allows us to enter longer data than GraphCodeBERT, since for the latter we have to count the space for data flow information. 

\paragraph{Inputs}
CodeBERT is able to process both source code and natural language, \eg comments and documentation. In our case, we did not exploit the possibility of using comments, because the input length is limited. 
The input passed to CodeBERT is the concatenation of two segments with a special separator token, namely: 
\begin{center}
 \([CLS], w1, w2, ... wn, [SEP], c1, c2, ..., cm, [EOS]. \)
\end{center}
Where \textit{Wi} is a sequence of natural language text, and \textit{Ci} is a sequence of code elements. [CLS] is a special token placed in front of the two segments, whose final hidden representation is considered as the representation of the whole sequence, used for the classification.
To make the source code of a test case match these expected inputs, we start by filtering it from extra spaces such as line breaks and tabs. Then, we tokenize the sequence and add the special tokens: CLS at the beginning and SEP at the end. We convert the sequence of tokens into IDs as each token is assigned an ID that corresponds to its index in the vocabulary file of CodeBERT. This sequence is passed to the CodeBERT model, which returns a vector representation. 

Figure~\ref{fig:using_codebert} illustrates this process.

\paragraph{Outputs}

CodeBERT output includes two representations, the first contains a contextual vector representation of each token in the sequence, and the second is the CLS representation having a size of 768, which is used to represent the whole sequence.
In the case of FlakyCat, we are interested in the CLS representation of the test code, which is used as a vector representation of the test cases. 

%\subsubsection{How did we use CodeBERT?}

\begin{figure*}[htbp]
\centerline{\includegraphics[width = 0.7\textwidth, scale=1]{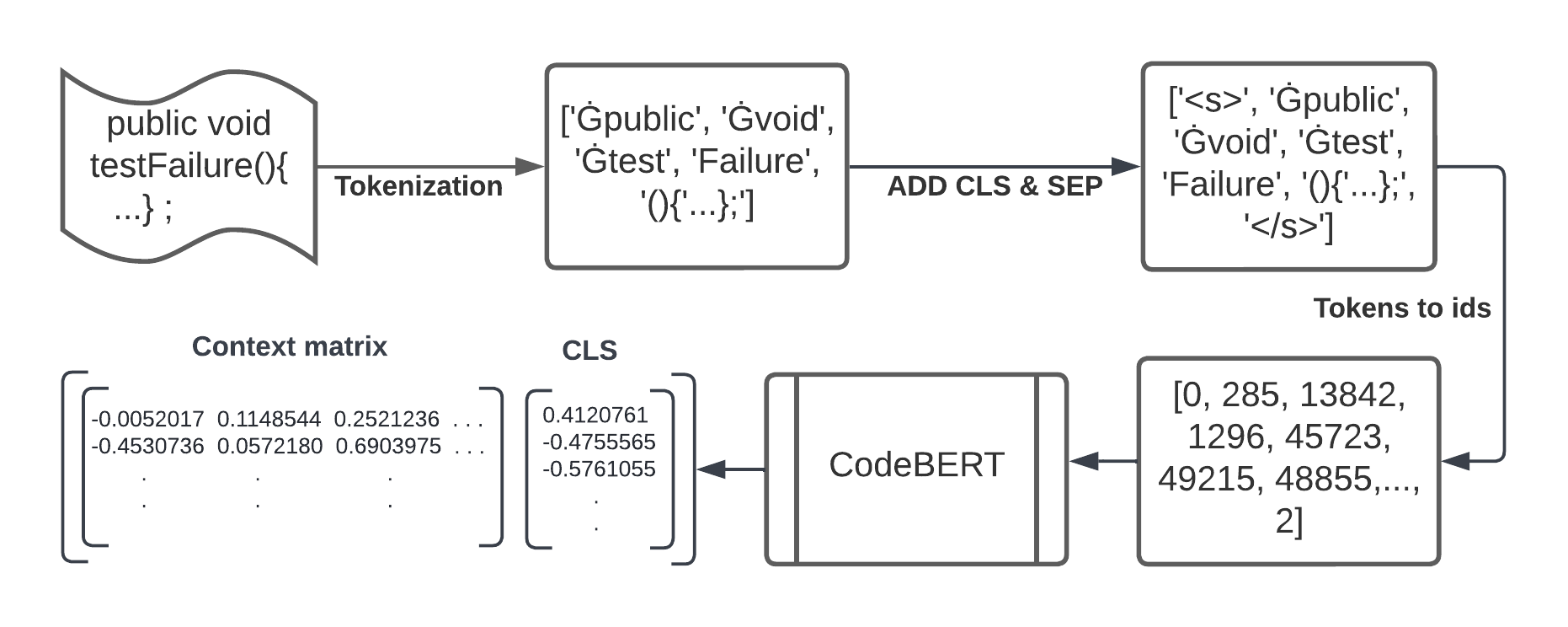}}
\caption{The process of converting the code source of a test case to a vector using CodeBERT, going through tokenization, then converting to IDs and applying the CodeBERT model to get the CLS vector}
\label{fig:using_codebert}
\end{figure*}

% Finally, the sequence is passed to the CodeBERT model, which returns the CLS representation, and this is what we use as input to our classifier.

%\subsubsection{Input limit}
%The maximum length of the codeBERT input sequence is 512, so any token exceeding this size will be discarded, and instructions important to classification can be deleted. Flakify \cite{fatima2021flakify} have addressed for this problem by only keeping statements related to test smells, but this eliminates statements that are relevant to flakiness categories such as time and network, but are not related to test smells. When this method has been applied, we notice that the majority of the conserved instructions are assertions. As a solution to this problem, we split long tests with a window of a size equal to the average length of all our test cases, which is 230 tokens, and we advance with a relatively small step (100 tokens) to include the flaky code in several parts. We classify each part and take the predicted labels with high confidence and select the most repeated label. 

\subsection{Step 2: Test categorization}

Unlike traditional machine learning classifiers that attempt to learn how to match an input $x$ to a probability $y$ by training the model in a large training dataset and then generalizing to unseen examples, Few-Shot learning (FSL) classifiers learn what makes the elements similar or belonging to the same class from only a few data. Facing the scarcity of data about flaky tests in general, selecting a Few-Shot classifier seems then to be a promising choice. A flaky test can be part of the category \textit{Async wait} just because it uses only one statement of explicit wait, FSL can focus on this statement and learn that it's more important than the others for the classification of this category. 
%A standard classifier will learn to classify the set of vocabulary and a single statement may be not important compared to the set, but an FSL classifier learns that this statement is common in the tests belonging to this category. 

In FSL, we call the item we want to classify a \textit{query}, and the \textit{support set} is a small set of data containing few examples for each class used to help the model to make classifications based on similarity. 
To classify flaky tests according to their flakiness category, we compute the similarity between the query and all examples of each flakiness category in our Support Set and assign the label having the maximum similarity with the query. This classification is obviously performed in a space where all elements of the same class are similar or close to each other. This is achieved by a model called \textit{Siamese network}. Its task is to transform the data and project it into a space where all the elements of a same class are close to each other, and then we can simply classify the elements by computing their similarity.

\subsubsection{Vector transformation with Siamese networks}

\begin{figure}[htbp]
\centerline{\includegraphics[width = 0.5\textwidth, scale=1]{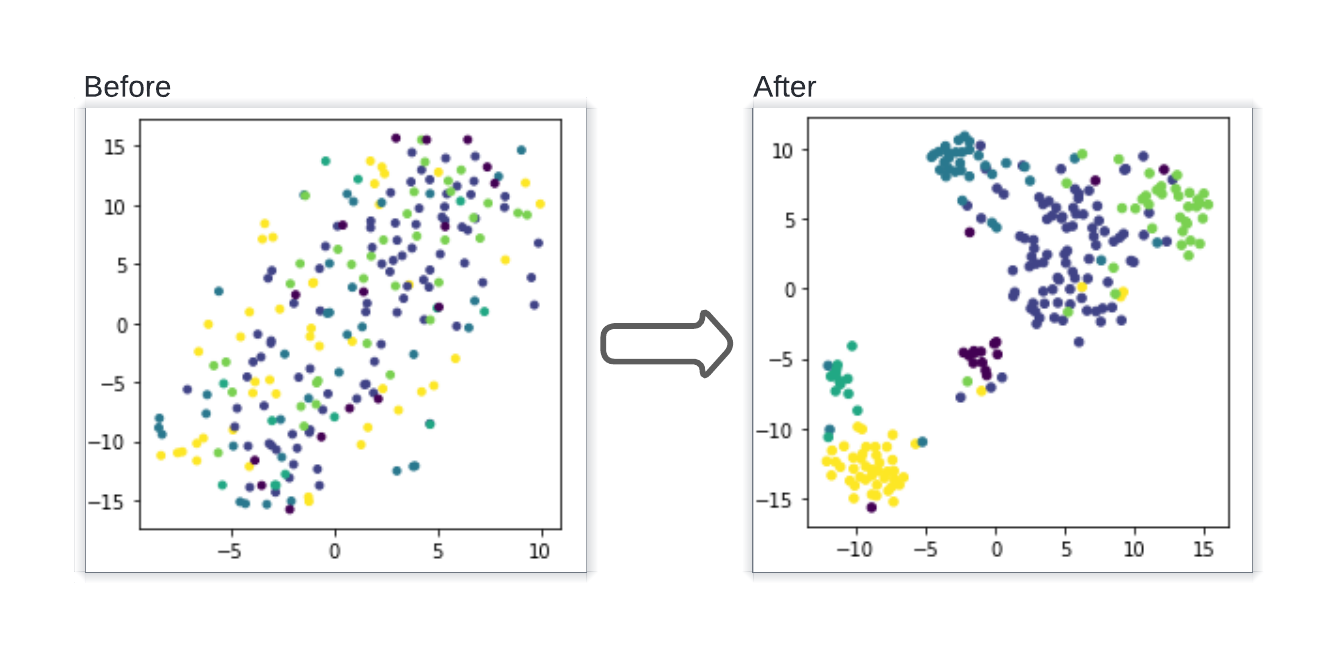}}
\caption{Visualization of data before and after training of the Siamese network with the triplet loss, which brings together the elements of the same class }
\label{fig:before_after}
\end{figure}

The Siamese network has knowledge of the similarity of elements of the same class. It processes two vectors in input and applies transformations that allow minimizing the distance between the two vectors if they share similar characteristics. Figure~\ref{fig:before_after} shows an example of the visualization of flaky test vectors before and after the Siamese network is applied. Since CodeBERT has no knowledge of the characteristics of flaky tests and only generates a general representation of the source code, the vectors produced are all similar.
However, the Siamese network learns which characteristics in these vectors are shared by tests of the same class, and thus allows to project vectors into a space that groups tests of the same flakiness category. After this step, it becomes possible to classify them with a simple similarity computation.

\paragraph{Siamese networks training}

\begin{figure}[htbp]
\centerline{\includegraphics[width = 0.3\textwidth]{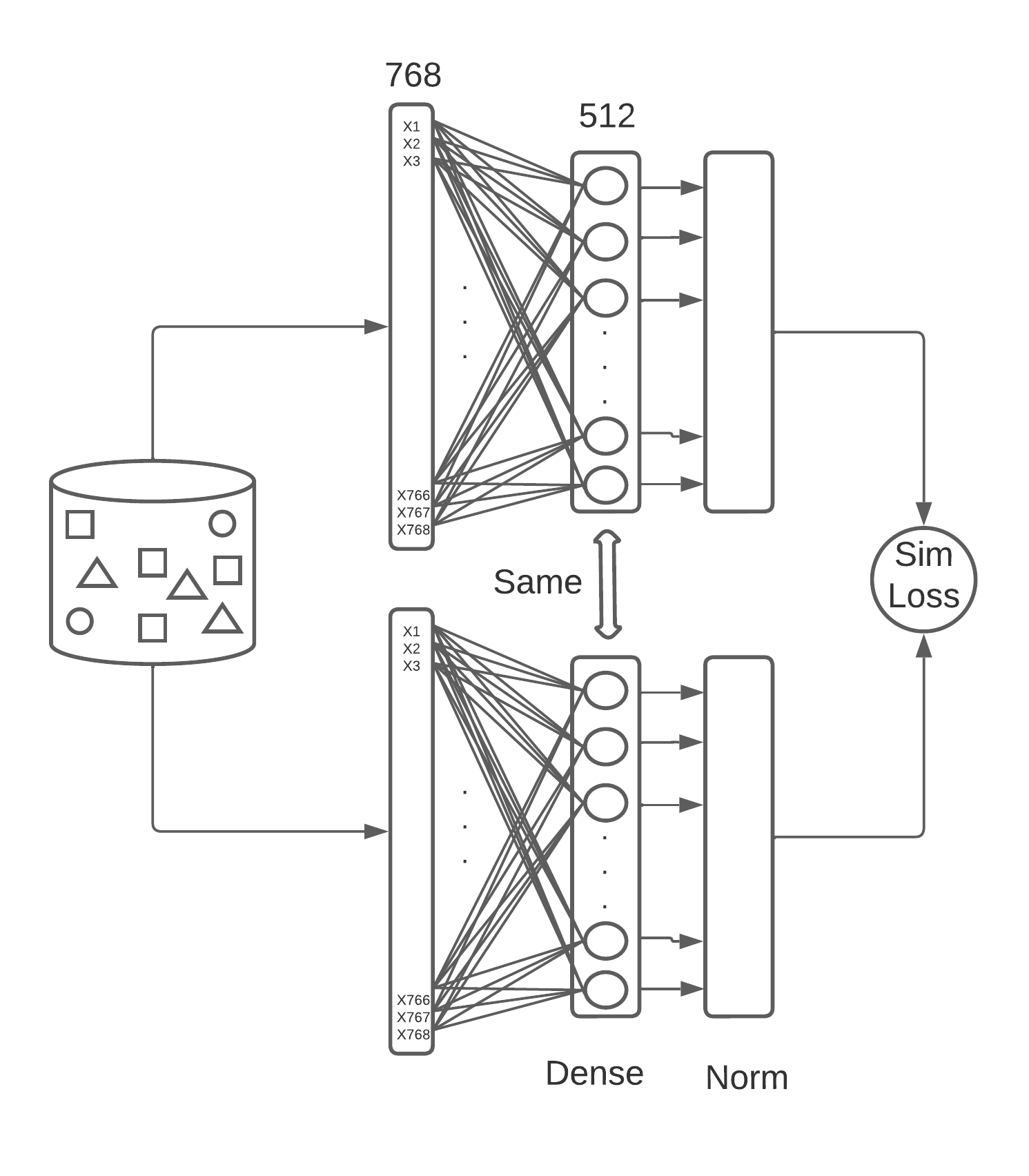}}
\caption{The architecture of the Siamese network used for our model and trained using the triplet Loss function }
\label{fig:Siamese_architetcure}
\end{figure}

Siamese networks have two identical sub-networks, each sub-network processes the input vector and performs transformations.
Both sub-networks are trained by calculating the similarity between the two inputs and using the similarity difference as a loss function.
Accordingly, the weights are adjusted to have a high similarity if the inputs belong to the same class. For the architecture of the sub-networks, we used a dense layer of 512 neurons and a normalization layer as shown in Figure~\ref{fig:Siamese_architetcure}.
We also performed a linear transformation to keep relations learnt by CodeBERT using attention mechanism introduced in the transformer architecture \cite{attention}. 
This model is trained using a Triplet Loss function, based on the calculation of similarity difference.

Let the Anchor $A$ be the reference input (it can be any input), the positive example $P$ is an input that has the same class as the Anchor, the negative example $N$ is an input that has a different class than the Anchor, $s()$ is the cosine similarity function, and $m$ is a fixed margin. The idea behind the Triplet Loss function is that we maximize the similarity between $A$ and $P$, and minimize the similarity between $A$ and $N$, so ideally $s(A, P)$ is large and $s(A, N)$ is small. The formula for this loss function is: 

\begin{center}
\( Loss = max(s(A, N) - s(A, P) + m, 0) \)
\end{center}
$m$ is an additional margin as we do not want $s(A, P)$ to be very close to $s(A, N)$, which would lead to a zero loss. 

To train the Siamese network with the triplet loss, we give as input batches of pairs with the same classes, and any other pair of a different class can be used as a negative example. We select the closest negative example to the anchor, such as $s(A,  N)$ $\simeq$ $s(A, P)$, which generates the largest loss and constitutes a challenge for the model learning. 

% \subsubsection{Classification of the siamese outputs}
 
% %The general process of classifying the category of flakiness is shown in Figure \ref{fig:general_acrh}. 
% We select a support set where each class is represented by few examples. For our study, we select the most centered examples to represent each class. 
% We start by transforming the source code of the tests into vectors by CodeBERT. For each couple (Query, element from t) we transform the embeddings by the Siamese network which brings together the vectors if they are of the same flakiness category and distances those of different categories, then we compute the cosine similarity between the two. We assign to the query the label of the test which maximizes the similarity score. 

\section{Interpretability technique}
\label{sec:inter}
% What if interpretability
Model interpretability refers to one's ability to interpret the decisions, recommendations, or in our case the predictions, of a model. Interpretability is a crucial step to increase trust in using a machine learning model. Indeed, it allows model creators to investigate potential biases in the learning processes and better assess the overall performance of their models.
On top of that, providing users, who are the developers in this case, with information about how the model came to its prediction can enhance the model adoption~\cite{carvalho2019machine}. 
%Interpretability is a growing field in AI as ML-based models are often viewed as black boxes.

% Existing techniques
Flakiness prediction approaches often rely on Information Gain to explain what features in the model yield the most information~\cite{Pinto2020,FlakeFlagger,camara2021use}. In the case of tree models, the reported information gain is given by the Gini importance (also known as Mean Decrease in Impurity)~\cite{Featurei12:online}.
Parry \etal~\cite{flake16} used SHapley Additive explanations (SHAP), which is another popular technique for model interpretability~\cite{shaponline}. 

% Use of CodeBERT increasing
As FlakyCat uses the CodeBERT representation of tests as input, using the previously mentioned techniques would not give understandable features. Thus, we decide to introduce a new technique to interpret CodeBERT-based models. We aim at understanding what information is learned by FlakyCat. 

% Our technique
Our technique is inspired by delta debugging algorithms. Delta debugging can help to isolate failing unit tests~\cite{zeller2002simplifying}. In our case, we are interested in code statements linked with the most influential information for the model's decision. 
To achieve this, 
%our idea is to train FlakyCat and check what statements influence the model's decision the most when removed from a test. 
%More precisely, 
we classify test cases and select only the ones that were correctly predicted (TP) as they contain information that was useful in the model's decision. We create new versions of each test.
Each version is a copy of the original test minus one statement that was removed. 
%We split statements in the test code based on the semicolon (statement separator token in Java). 
Next, we feed the new versions to FlakyCat. Among all new versions for one test, we keep the one for which the similarity score with the correct category endured the biggest drop. 
We consider the statement removed in this version as the most influential one.
When flakiness categorisation results are presented to developers, the most influential statement can be highlighted to explain the rationale behind the assigned category.
We can also envision scenarios where multiple influential statements are highlighted gradually based on their contribution to the classification result.

%The statement removed from this test is added to the list of textit{influential statements}. 

\section{Evaluation}
\label{sec:eval}
In this section, we explain our evaluation setting for FlakyCat.
First, we describe our data curation process, then, we present our approach for answering each of the three research questions.

\subsection{Data curation}
\label{dataset}

\subsubsection{Collection}
For our study, we had to collect a set of flaky tests containing their source code and their flakiness category.
We focused our collection efforts on one programming language, as training a classifier using code and tokens from different programming languages is more challenging. 
For the language choice, we opted for Java, which is the most common language in existing flakiness datasets. 
Nevertheless, as existing sets do not contain enough data about flakiness categories, we also built a new set of flaky tests from GitHub that we classify manually.
%and the language most popular language used in previous studies of flakiness, we opted for it.
%To build our dataset we use two main sources: flaky tests already classified in previous studies, and a new set of flaky tests from GitHub that we classify manually.

\paragraph{Existing datasets} 
There is no large public dataset of flaky tests labelled according to their category of flakiness. Most of the existing data are split into flaky and non-flaky tests and are used for binary classification such as FlakeFlagger~\cite{FlakeFlagger} and DeFlaker~\cite{deflaker}. There is also the Illinois' dataset\footnote{https://mir.cs.illinois.edu/flakytests/}, which is partially classified into order-dependent and implementation-dependent flaky tests. Regarding the data classified by flakiness categories defined by Luo~\etal \cite{Luo2014} and Eck~\etal~\cite{eck}, there is only limited data available used for analysis in previous empirical studies about flakiness. Luo~\etal~\cite{Luo2014} analyzed 200 commits and classified 135 commits into 10 flakiness categories. Using the same taxonomy, Eck~\etal~\cite{eck} classified 200 tests into 10 categories, including four new categories of flakiness. 
%For their study on flakiness in Python, %Gruber~\etal~\cite{Gruber2021} classified 100 flaky tests written in Python according to their flakiness categories. 
We retrieved 135 classified tests from the dataset of Luo~\etal, however, we were not able to access the set of Eck~\etal. 
We also identified a new dataset of classified flaky tests\footnote{https://github.com/Test-Flaky/TSE22}, from which we recovered 114 tests. 

\paragraph{New dataset}
To expand our dataset, we explore GitHub projects and search for flakiness-fixing commits that mention a flakiness category in their messages.
%Such commits allow us to identify 
%, we look for commits in Java projects on GitHub that solve a problem of a flaky test by looking for 
In this search, we use flakiness-related keywords such as \textit{Flaky} and \textit{Intermit} in the commit messages. 
To ensure that the commit refers to a flakiness category, 
%Since this selection yields a large number of commits that would take too long to inspect one by one, further refinement was required. We searched for other 
we further filter commits by specific keywords related to each category: \textit{thread, concurrence, deadlock, race condition} for Concurrency, \textit{time, hour, seconds, date format, local date} for Time, \textit{port, server, network, http, socket} for Network and \textit{rand} for Random.
After the search, we rely on the developers' explanation in the commit message and on the provided fix to classify tests into the different flakiness categories listed in the literature.
This collection allowed us to obtain 214 categorized tests. To ensure the correctness of our manual classification, the first two authors of the paper performed a double-check on the whole dataset to identify.

\subsubsection{Filtering}
The previous step allowed us to collect 214 categorized flaky tests.
In this step, we filter out tests that are not adequate for our study.
% As already mentioned, we only take test cases having the flakiness problem in the test code. Some tests having the problem in the production code or config files have been ignored. 
In particular, some data points in the existing datasets were missing attributes necessary for data extraction, such as the name of the test method or the download link.
As we were unable to obtain the missing fields, we filtered out these points.

The filtering reduced the number of test cases from 135 to 79 for the dataset of Luo~\etal, and from 114 to 50 in the new identified dataset. For the data we collected ourselves, we accounted for the filters from the beginning, so we retain the number of 214 successfully collected and extracted test cases, leaving us with a set of 343 categorized flaky tests. 

\subsubsection{Processing}
After filling all the necessary attributes: the test case name, flakiness category, test file name, and project URL, we download the code files and extract test methods using the spoon library\footnote{https://github.com/INRIA/spoon}. At this stage, all comments have been deleted from the source code to restrict CodeBERT to code statements.

\subsubsection{Data selection}

The final dataset contains 343 flaky tests distributed over 10 flakiness categories. 
Table~\ref{tabData} illustrates this repartition.

The collected flaky tests are not distributed evenly across categories of flakiness. Just as shown in past empirical studies~\cite{Luo2014,Gruber2021}, some categories, such as \textit{Async waits}, are more prevalent than others. 
Our approach uses Few-Shot learning to learn from limited datasets. 
Still, it requires a certain amount of examples to learn common patterns from each category. 
We decided to have at least 30 tests in a category to consider it. This number is commonly accepted by statisticians as a threshold to have representativeness~\cite{why30}. 
Also, we decided to discard the category of \textit{Test order dependency} as it is a particular case where the flakiness is specific to the order of the test suite. This leaves us with four categories, highlighted in grey in the table: \textit{Async waits}, \textit{Unordered collections}, \textit{Concurrency}, and \textit{Time}.

\subsubsection{Data augmentation}
Facing the challenge of learning from few data, we over-sampled our dataset in a similar way to SMOTE~\cite{chawla2002smote}. We duplicated tests by mutating only the code elements that have no influence on flakiness. This includes variable names and constants such as strings, numbers and booleans. We used the Spoon library for the detection of these elements, and we replaced them with randomly generated significant words. As a result, the total number of tests after data augmentation is 639.

\begin{table}[htbp]
\centering
\caption{Final dataset. The Highlighted rows are the data used to train and test the model. The original data refers to the data we collected, and the augmented data is the additional data we created after oversampling}
\begin{tabular}{|c|c|c|}
\hline
\textbf{Class}&\multicolumn{2}{|c|}{\textbf{Data}} \\
\cline{2-3} 
\textbf{} & \textbf{\textit{Original}}& \textbf{\textit{Augmented}} \\
\hline
	
\rowcolor{Gray}
Async waits & 89 & 285 \\
\hline
\rowcolor{Gray}
Unordered collections & 45 & 136\\
\hline
\rowcolor{Gray}
Concurrency & 36 & 113 \\
\hline
\rowcolor{Gray}
Time & 35 & 105 \\
\hline
Test order dependency & 33 & 99 \\ 
\hline
Network & 16 & 51 \\
\hline
Randomness & 13 & 40 \\
\hline
Test case timeout & 9 & 29 \\
\hline 
Resource leak & 5 & 17 \\
\hline
Platform dependency & 2 & 7 \\
\hline 
Too restrictive range & 2 & 7 \\
\hline 
I/O & 2 & 6 \\
\hline 
\end{tabular}
\label{tabData}
\end{table}

\subsection{Experimental design}
\subsubsection{Baseline}
~~ \\
The automatic classification of flaky tests according to their category is a task that has not already been performed automatically.
Previous studies analyzing the categories of flakiness relied on a manual classification of tests. 
Hence, as a baseline, we compare FlakyCat with existing approaches used to classify tests as flaky or not, including the vocabulary-based approach \cite{pinto2020vocabulary}, and the smell-based approach \cite{camara2021use}. Our motivation is to determine whether it is possible to make this classification based on limited data, and which classifier and code representation are the most suitable for flakiness classification. 

More specifically, we compare our FSL-based approach with traditional classifiers from the Scikit-learn library~\cite{pedregosa2011scikit} used by previous studies on flakiness prediction~\cite{pinto2020vocabulary,camara2021use,Camara2021VocabExtendedReplication}: Random Forest (RF), Support Vector Machine (SVM), Decision Tree (DT), K-Nearest Neighbour (KNN). 
%We cannot use Naive Bayes because the CodeBERT vectors include negative values. 
We also compare the code representation used in our classification approach with the different source code representations used in the context of flaky tests, in particular test smells and vocabulary. 

For the classification based on test smells, we use the 21 smells detected by tsDetect \cite{tsdetect}, to generate vectors indicating the presence of each smell detected by the tool, in the same way as in the study of Camara \etal \cite{camara2021use}. As for the vocabulary-based classification, we use token occurrence vectors, as in the article by Pinto \etal \cite{pinto2020vocabulary}. We tokenize the code and apply standard pre-processing like stemming, then calculate occurrences of each token.

We use a 4-fold stratified cross-validation to assess the predictive performances of our model. FlakyCat relies on a Siamese network. It is trained with combinations of data by indicating whether these data are similar or not so that the model can learn what makes them similar. Since we train with combined data, the balancing of data is not required, because it is automatically over-sampled.

As the augmented samples in our dataset are very similar to the original ones, it was important to keep them in the same sets for training, so that no augmented samples are leaked in the test set. For the support set used for classification, we select the most centred examples to represent each class.

\subsubsection{Parameters}
~~ \\
For FlakyCat, we select the parameters by testing combinations of the most important parameters that have a direct impact on the model performance, which include the similarity margin used in the triplet loss function and the number of data pairs used to train the Siamese model. As shown in Figure~\ref{fig:parameters}, we choose a margin of 0.3 and we fix the number of data pairs at 10,000. 
We select the standard value $lr = 0.01$ for the learning rate. The change of this value had no impact on the results.
For baseline classifiers, we keep the standard values used by previous works. We varied the number of trees in the Random Forest classifier, we tested values from 100 to 1000 with a step of 100.
We observed that this does not make much difference regarding the F1 score (<1\%), and we identified 200 as the number giving the best results. 

\subsubsection{Evaluation metrics}
~~ \\
We use the standard evaluation metrics to compare between classifiers, including precision, recall, Matthews correlation coefficient (MCC), F1-score, and Area under the ROC curve (AUC).
These metrics have been used to evaluate the performance of classifiers, including binary classification of flaky tests~\cite{pinto2020vocabulary,camara2021use,fatima2021flakify}.

% margin , nb steps , number support set , learning rate , 

\begin{figure}[htbp]
\centerline{\includegraphics[scale=0.7]{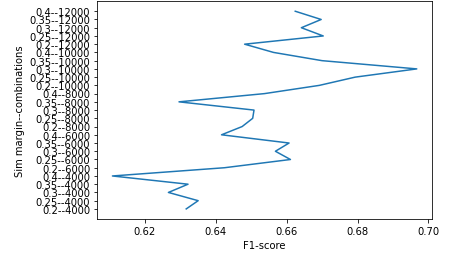}}
\caption{F1-score for different values of similarity margin and combinations}
\label{fig:parameters}
\end{figure}

\subsection{RQ1: How effective is FlakyCat compared to approaches
based on traditional supervised learning?}
This question aims to evaluate FlakyCat and compare it to relevant baselines.
In the following, we explain the choice of these baselines and our selection of parameters and evaluation metrics.

\subsection{RQ2: How effective is FlakyCat in predicting each one
of the considered flakiness categories?}
This question aims to evaluate FlakyCat's ability to classify the different categories of flakiness.
To perform this, we split the dataset into four sets following the categories: \textit{Async waits}, \textit{Unordered collections}, \textit{Concurrency}, and \textit{Time}.
Then, we use the same settings as for RQ1 to tune the Siamese network, train it, and evaluate it for each category.

\subsection{RQ3: How do statements of the test code influence the
predictions of FlakyCat?}

We applied the technique we proposed in section~\ref{sec:inter} for CodeBERT-based model interpretability to FlakyCat. 
%We collected 119 statements of interest, important for the decision-making of FlakyCat.
We collected the most influential statement for each flaky test correctly classified. 
We then proceed to investigate those statements. First, we regroup statements by category of flakiness (according to the test they belong to). Then, we want to give details on what type of statements FlakyCat found useful. To do so, we look through the list of statements and attempt to identify recurring code statements and categorise them. The process of identifying statement types is subjective and inspired by qualitative research. The idea is to find statement types of interest for our specific case of flakiness classification. We identified 8 types of statements that are related to some aspects of flakiness in each category:
\begin{itemize}
 \item \textbf{Control flow:} Groups statements of loops and conditions.
 \item \textbf{Asserts:} All types of assertions in tests. 
 \item \textbf{Threads:} All thread control statements. 
 \item \textbf{Constants:} Constant values of numbers, strings, and booleans. 
 \item \textbf{Waits:} All explicit wait statement. 
 \item \textbf{Time-related:} Statements that perform operations on time values, dates. 
 \item \textbf{External/API calls:} Groups network / API calls, and manipulation of external resources such as files and databases.
 \item \textbf{New instances:} Statements creating new instances or objects.
\end{itemize}
In this question, we investigate the prevalence of these statement types in each flakiness category.

\section{Results}
\label{sec:results}

\subsection{RQ1: How effective is FlakyCat compared to approaches
based on traditional supervised learning? }

Following the outlined experimental design, we trained and tested FlakyCat and the four traditional classifiers, using the three source code representations, the vectors obtained from CodeBERT, the vectors based on vocabulary, and the ones on tests smells. The obtained results are presented in Table~\ref{scores}. The results show that FlakyCat achieves the best performance for all evaluation metrics. 
It obtained an average weighted F1 score of 70\% and a precision of 71\%.
We get an MCC of 0.58 (bounds for this metric are between -1 and 1), being close to 1 means a perfect classification. We have an AUC of 0.78, which shows that the model is highly distinguishing between classes. 

\paragraph{Representation effect}
Regarding the three code representations, CodeBERT achieves the best performance among all classifiers, with an F1 score between 0.45 and 0.7 for the five classifiers. 
By using the vocabulary vectors, all classifiers do not exceed an F1 score of 0.6.
The representation based on test smells yields the worst result, with the best F1 score being 0.26.
This means that CodeBERT is able to learn more information about the code and the flaky test behaviours.

\paragraph{Classifier effect}
Regarding the choice of classifier, we find that the Few-Shot classifier based on similarity achieves the best performance using the representations based on CodeBERT and vocabulary. Among traditional classifiers, Random Forest obtains the best results, as reported in previous flaky test classification studies~\cite{Pinto2020,Haben2021}. While for the smells-based representation, none of the classifiers showed promising results. Using this code representation, the SVM classifier turned out to achieve the best F1 score: 0.26. We conclude that the information present in the test smells was not helpful to learn how to predict flaky tests category.  \\

\begin{table*}[htbp]
\caption{Comparing performances of FlakyCat (CodeBERT and Few-Shot Learning) with traditional machine learning classifiers}
\begin{center}
\begin{tabular}{|c|c|c|c|c|c|c|c|c|c|c|c|c|c|c|c|}
\hline
\textbf{Model} & \multicolumn{5}{|c|}{\textbf{Smells-based}} & \multicolumn{5}{|c|}{\textbf{Vocabulary-based}} &\multicolumn{5}{|c|}{\textbf{CodeBERT-based}} \\
\cline{2-16} 
\textbf{} & \textbf{\textit{Precision}}& \textbf{\textit{Recall}} & \textbf{\textit{ MCC}}& \textbf{\textit{F1}} & \textbf{\textit{AUC}}& \textbf{\textit{Precision}}& \textbf{\textit{Recall}} & \textbf{\textit{ MCC}}& \textbf{\textit{F1}} & \textbf{\textit{AUC}} & \textbf{\textit{Precision}}& \textbf{\textit{Recall}} &\textbf{\textit{ MCC}}& \textbf{\textit{F1}} & \textbf{\textit{AUC}}\\
\hline
SVM & 0.19 & 0.44 & 0.00 & 0.26 & 0.50 & 0.48 & 0.47 & 0.18 & 0.35 & 0.54 & 0.19 & 0.43 & 0.00 & 0.26 & 0.50 \\
\hline
KNN & 0.10 & 0.2&0.0 1 & 0.10 & 0.51 & 0.44 & 0.43 & 0.12 & 0.37 & 0.55 & 0.50 & 0.50 & 0.23 & 0.45 & 0.59 \\
\hline
DT & 0.17 & 0.36 & -0.07 & 0.22 & 0.46 & 0.43 & 0.47 & 0.24 & 0.45 & 0.62 & 0.51 & 0.49 & 0.28 & 0.49 & 0.64 \\
\hline
RF & 0.15 & 0.31 & -0.08 & 0.20 & 0.45 & 0.62 & 0.6 & 0.42 & 0.53 & 0.66 & 0.63 & 0.63 & 0.46 & 0.59 & 0.70 \\
\hline
\textbf{FSL} & 0.11 & 0.31 & -0.01 & 0.16 & 0.50 & 0.63 & 0.63 & 0.46 & 0.60 & 0.71 & \textbf{0.71} &\textbf{0.70} & \textbf{0.58} & \textbf{0.70} & \textbf{0.78} \\
\hline
\end{tabular}
\label{scores}
\end{center}
\end{table*}

\begin{tcolorbox}

Overall, our results show that it is possible to automatically classify flaky test categories with limited dataset size. CodeBERT is the best approach to represent flaky test source code and Few-Shot learning performs better than traditional machine learning classifiers. 
\end{tcolorbox}

\subsection{RQ2: How effective is FlakyCat in predicting each one
of the considered flakiness categories?}

Table~\ref{tab_classes} shows performances achieved by FlakyCat for each of the four flakiness categories that we have selected. Figure~\ref{fig:FSL}  shows that the category \textit{Unordered collections} is the easiest for the model to classify, with a precision of 0.84 and an F1 score of 0.85 (average across folds). The category \textit{Async waits} and \textit{Time} respectively have a precision of 0.73 and 0.65. \textit{Concurrency} performances are lower with a precision of 0.48. 
We suspect that concurrency issues happen in many cases in the code under test. As FlakyCat only relies on the test source code, this would indeed explain why performances are lower in this case. Another supposition is that concurrency issues and asynchronous waits are sometimes closely related. For example, a thread incorrectly waiting for another one to finish might be considered both a concurrency issue and an asynchronous wait. \\

\begin{figure}[htbp]
\centerline{\includegraphics[scale=0.7]{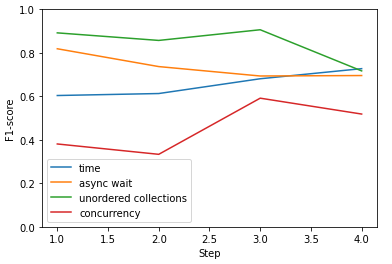}}
\caption{F-score per flakiness category using FlakyCat}
\label{fig:FSL}
\end{figure}

\begin{table}[htbp]
\caption{Performances by categories using Few-Shot classifier}
\begin{center}
\begin{tabular}{|c|c|c|c|}
\hline

% the scores avg for the cross validation 

\textbf{Category} & \textbf{\textit{Precision}}& \textbf{\textit{ Recall}} & \textbf{\textit{Weighted F1-score}} \\
\hline
Async waits & 0.73 & 0.75 & 0.74 \\
\hline
Concurrency & 0.48 & 0.48 & 0.46 \\
\hline
Time & 0.65 & 0.68 & 0.66 \\
\hline
Unordered collections & 0.84 & 0.80 & 0.85 \\

\hline
\end{tabular}
\label{tab_classes}
\end{center}
\end{table}

\begin{tcolorbox}

While the three flakiness categories \textit{Unordered collections}, \textit{Async waits} and \textit{Time} show good ability to be detected automatically, \textit{Concurrency} remains difficult to detect by relying only on the test case code. 

\end{tcolorbox}

\subsection{RQ3: How do statements of the test code influence the
predictions of FlakyCat?}

Table~\ref{TabStatments} reports the prevalence (\%) of the different types of statements among all influential statements per flakiness category, \eg 100\% Asserts in the time category would mean that all influential statements for the time category contain assert statements. 

The results in Table~\ref{TabStatments} show that the \textit{Control flow}, \textit{Constants}, and \textit{New instances} statements are almost evenly distributed. We conclude that those statement types are not correlated to the specificity of flakiness categories under study. Compared to other flakiness categories, the percentage of assertions in the influential statements of  \textit{Time} and \textit{Unordered collections} is high, 55\% and 50\% respectively. Based on our analysis, this includes in particular assertions that perform exact comparisons, such as \texttt{assertEquals()}, between constant values and collection items, or dates for example. 
43\% of influential statements in the \textit{Concurrency} category include some thread manipulation, and 17\% for the \textit{Async Waits} category, while the rest of the categories have none. 
Statements containing explicit waits represent respectively 17\% and 25\% for \textit{Async Waits} and \textit{Concurrency} categories, but zero for the others. Statements containing time values are most common in the \textit{Time} category with 85\%. We note that they appear as well in a small proportion, 12.8\% and 12.5\% respectively, for \textit{Async Waits} and \textit{Concurrency}. Statements from the external/API calls group are mainly found in the \textit{Async waits} and \textit{Concurrency} categories, this includes network calls and manipulation of external resources.

The results show that FlakyCat is able to differentiate between the features that are important to each flakiness category by considering the correlation between the types of statements and flakiness categories. This also suggests that CodeBERT is able to grasp some semantics from the test code. \\

\begin{tcolorbox}
Our analysis of the most influential statements shows that the statements influencing the predictions of FlakyCat are correlated to flakiness categories.
By highlighting these statements, our interpretability technique can help developers understand flaky tests and their categories.
\end{tcolorbox}

% asserts -> Time & unordered collections
% New instance & constants -> same distribution across categories
% time -> mostly time but also async waits and concurrency ( not for collections )
% threads & waits -> only async waits and concurrency
% external API calls -> mainly Async waits & concurrency

% 1. model learnt what needs to be learnt and Codebert learns good features 
% 2. 

\begin{table*}[htbp]
\caption{Prevalence of the different types of statements in each flakiness category}
\begin{center}
\resizebox{\textwidth}{!}{
\begin{tabular}{|c|c|c|c|c|c|c|c|c|c|}
\hline
& NB statements & Control flow &	Asserts &	Threads &	Constants &	Waits &	Time Related	& External/API calls &	New instance \\
\hline 

Async Waits & 47 & 2.1\% & 17.0 \% & 17.0 \% & 66.0 \% & 17.0 \% & 12.8\% & 23.4\% & 8.5\% \\

\hline 
Concurrency & 16 & 0,0\% &	6,3\% &	43,8\% &	56,3\% &	25,0\% &	12,5\% &	31,3\% &	12,5\% \\
\hline 
Time & 20 & 10,0\%	& 55,0\% &	0\% &	60,0\% &	0\% &	85,0\% &	5,0\% &	15,0\% \\
\hline 
Unordered collections & 36 & 8,3\% & 50,0\% &	0\% &	66,7\% &	0\% &	0\% &	8,3\% & 13,9\% \\
\hline
\end{tabular}
}
\end{center}
\label{TabStatments}
\end{table*}
\vspace{-4mm}

\section{Discussion}
\label{sec:discussion}
\subsection{The effect of adding additional categories}
Our results showed that flakiness categories can be classified automatically. We carried out our main experiments with four categories of flakiness for which we had a reasonable number of tests. Still, we believe that one interesting aspect of our study is understanding the impact of adding other categories to FlakyCat. 
For this, we investigate the performance of FlakyCat in each category (similarly to RQ2), but we consider two more categories, \textit{Network} and \textit{Random}.
These are the next two categories with the most samples in our dataset with 16 and 13 flaky tests, respectively. The F1 scores and the accuracy obtained for each category are presented in figure \ref{fig:fsl_add_class}. 

Compared to the results previously reported in Table~\ref{tab_classes}, we observe that the performances of each category are slightly impacted.
The \textit{Async waits} category is the most impacted one. 
Indeed, after adding two categories, we get an overall F1 score of 0.57, where the added categories get the worst results. 
This performance drop is caused by multiple factors.
The first one is the increase in the number of classes for the classifier from 4 to 6.
The discrimination between four classes is much easier than six.
Indeed, the top four categories become more difficult to distinguish, which means that the added categories have common characteristics with them. 
Secondly, the overall F1 score is affected by the poor performances observed in the new two categories.
These performances can be a result of the number of examples in these categories (less than 30), which does not allow the classifier to learn the similarities between tests in these categories, and even if it learns, the features are still not mineralizable. 

\begin{figure}[htbp]
\centerline{\includegraphics[scale=0.7]{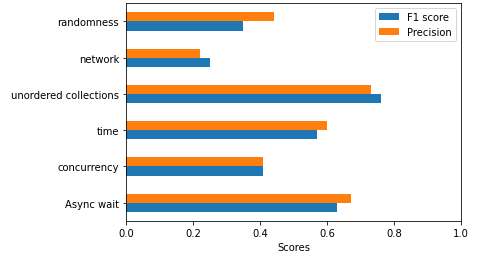}}
\caption{F-score and precision per category when considering six flakiness categories}
\label{fig:fsl_add_class}
\end{figure}

To further understand these results, we inspect a test that was misclassified by FlakyCat.
The test is presented in the listing~\ref{fig:test_example}\footnote{https://github.com/apache/hbase/commit/e89712d29dd91be4}.
Based on the dataset of Luo~\etal~\cite{Luo2014}, the original flakiness label of this test is \textit{Network}, because the test fails intermittently due to a lack of online dependency.
However, FlakyCat assigns it first to the category \textit{Concurrency}, then the second label is \textit{Async waits}, and the third one is \textit{Network}. Analyzing the whole test, we can notice that it uses multiple threads, unordered collections, and asynchronous wait for a fixed time, so the test may be flaky for other reasons.
Hence, statically multiple categories of flakiness can be assigned to it. In this case, the model can't make much of a distinction, because it works with similarities, and the test is similar to more than one category. 

\begin{figure}[htbp]
\centerline{\includegraphics[scale=0.62]{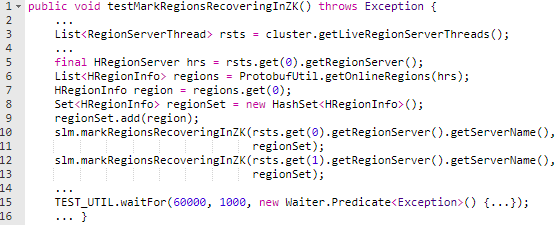}}
\caption{A flaky test from the Category \textit{Network} taken from the project HBase.}
\label{fig:test_example}
\end{figure}

\subsection{Correlation between the statements influencing FlakyCat and flakiness causes}
The results of RQ3 highlight the type of code statements on which our model relies to make predictions. Looking at these results, we can notice that, according to the model, \textit{Async waits} and \textit{Concurrency} categories have similar types of statements behind them.
These statements are related to the use of threads, wait statements with fixed time values, constants and external calls. This means that the two categories have common characteristics. Moreover, external calls, including network calls which constitute a separate flakiness category, are also frequent in the \textit{Async waits} and \textit{Concurrency} categories. In addition, time-related statements, which are mainly present in the \textit{Time} flakiness category, occur also in the \textit{Async waits} and \textit{Concurrency} categories. This implies that several defined flakiness categories can intersect (or be included in one another). The results may also draw our attention to the fact that even some elements of the code that may seem important when debugging flakiness problems may be part of the problem. As table~\ref{TabStatments} shows, constants and asserts are among the important elements for the classification of the categories \textit{Time} and \textit{Unordered collection}. This is because flaky tests that are caused by unordered collection try to assert that elements of an unordered collection are exactly equal to constant values with a fixed order.
The same issue occurs in the \textit{Time} category as tests compare time values with different precision.
This means that developers and tools should be careful with exact equality assertions to avoid these categories of flakiness.
 
Future research studies may further investigate the precise causes of flakiness and make clear distinctions between categories. The application of machine learning to determine the causes of flakiness is promising and should receive attention while considering that different categories of flakiness may be included in a single test.

% \paragraph{Limits}
% Another threat happens when we ask our model to classify a test from an unsupported category. In this case, the model can only give one of the category it was trained on. Still, we believe flakiness categories are not orthogonal and that it is possible for a flaky test to belong to several categories. For example, a flaky test relying on network or I/O can flake because of an asynchronous wait. Thus, we consider that the prediction of our model can still provide value to understand the root cause in those cases. 
% One limit of our approach is derived from the use of CodeBERT. To get its CodeBERT representation, the number of tokens in a test can't exceed 512. When collecting our data, tests were discarded when they were too long. We can't guarantee that our approach would generalize to longer and higher level tests. The same applies for other languages.

\section{Threats to validity}
\label{sec:threats}
\paragraph{Internal validity}
One threat to the internal validity is related to the dataset we used in our study. Flaky tests were gathered from different sources, as explained in section~\ref{dataset}. It is possible that flaky tests were assigned to the wrong label, which would impact the training and evaluation of our model. Certifying the category based on the test source code is complex and can as well be subjective. To ensure the quality of the data, the first two authors reviewed the collected flaky tests and confirmed their belonging to the assigned category.

Similarly, the identification of statement types in RQ3 required a manual analysis of the most influential statements. 
Hence, the identified types can be subjective and the assignment of statements is prone to human errors.
To mitigate this risk, we kept the statement types factual, \eg control flow and asserts.
This allows us to avoid assignment ambiguities and intersections between the different statement types.
\paragraph{External validity}
The first threat to external validity is the generalizability of our approach. In this study, we train a model to recognize flaky tests from four of the most prevalent categories, but we are not sure of the performances in other categories. We discussed the addition of two categories (Network and Randomness), and retrieved that the number of examples is one of the influencing factors. 
%but our approach might not perform as well as for the categories we considered. 

\paragraph{Construct validity}
One potential threat to construct validity regards the metrics used for the evaluation study. To alleviate this threat, we report MCC, F1-score, and AUC metrics in addition to the commonly-used precision and recall. As our data is not evenly distributed across the different categories, we report the weighted F1 score.

% Finally, regarding our technique to analyse decisive statements, a threat is that the importance might not rely in one statement only but in bigger context. 

\section{Conclusion}

Test flakiness is considered as a major issue in software testing as it disrupts CI pipelines and breaks trust in regression testing.
%crippling Continuous Integration and wasting developers time. 
Detecting flaky tests is resourceful as it can require many reruns to reproduce failures. To facilitate the detection, more and more studies suggest static and dynamic approaches to predict if a test is flaky or not.
However, detecting flaky tests constitutes only a part of the challenge since it remains difficult for developers to understand the root causes of flakiness. Such understanding is vital for addressing the problem, \ie fixing the cause of flakiness. At the same time, researchers would gain more insights based on this information. So far, only a few automated fixing approaches were suggested and these are focusing on one category of flakiness. Knowing the category of flakiness for a given flaky test is thus a key information. 

With our work, we propose a new approach to this problem that aims at classifying previously identified flaky tests in their corresponding category. We present FlakyCat, a Siamese network-based multi-class classifier that relies on CodeBERT's code representation. FlakyCat addresses the problem of data scarcity in the field of flakiness by leveraging the Few-Shot learning capabilities of Siamese networks to allow the learning of flakiness categories from small sets of flaky tests. As part of our evaluation of FlakyCat, we collect and make available a dataset of 343 flaky tests with information about their category of flakiness. 

Our empirical evaluation shows that FlakyCat performs the best compared to other code representations and traditional classification models used by previous flakiness prediction studies. In particular, we reach a weighted F1 score of 70\%. We also analysed the performances with respect to each category of flakiness. We found that flaky tests belonging to \textit{Async waits}, \textit{Unordered collections} and \textit{Time} are the easiest to classify, whereas flaky tests from the \textit{Concurrency} category are more challenging to predict. Finally, we present a new technique to explain CodeBERT-based machine learning models which is inspired by delta-debugging. This technique helps in explaining what code elements are learnt by models and could give useful information to developers who wish to understand flakiness's root causes. 

% \section*{Acknowledgment}

\bibliographystyle{IEEEtran}
\bibliography{main}

\end{document}